\documentclass[aps,prd,preprintnumbers,showpacs,showkeys,nofootinbib,
superscriptaddress,fleqn,floatfix,tightenlines,10pt]{revtex4-1}
\usepackage{amsmath,amsfonts,amssymb,amscd,amsxtra,amsthm}
\usepackage{graphicx}  
\usepackage{epstopdf}
\usepackage{dcolumn}  
\usepackage{bm}          
\usepackage{slashed}
\usepackage{cancel}
\usepackage{float} 
\usepackage{mathtools}
\usepackage{amsbsy}
\usepackage{amstext}

\providecommand{\tabularnewline}{\\}
\usepackage[utf8]{inputenc} 
\usepackage{booktabs} 
\usepackage[normalem]{ulem} 
\usepackage[dvipsnames]{xcolor} 
\usepackage{tabularx}
\usepackage{enumitem}  
\usepackage{array} 
\usepackage{slashed}
\usepackage{tikz}
\usepackage{float}
\usepackage{multirow}
\renewcommand\sout{\bgroup \color{red} \ULdepth=-.5ex \ULset}

\makeatletter

\begin{document}  
\preprint{INHA-NTG-09/2019}
\title{Magnetic transitions and radiative decays of singly heavy baryons}
\author{Ghil-Seok Yang}
\email[E-mail: ]{ghsyang@ssu.ac.kr}
\affiliation{Department of Physics, Soongsil University, Seoul 06978,
Republic of Korea}

\author{Hyun-Chul Kim}
\email[E-mail: ]{hchkim@inha.ac.kr}
\affiliation{Department of Physics, Inha University, Incheon 22212,
Republic of Korea}
\affiliation{School of Physics, Korea Institute for Advanced Study 
  (KIAS), Seoul 02455, Republic of Korea}
\date{\today}
\begin{abstract}
A pion mean-field approach allows one to investigate light and singly
heavy baryons on an equal footing. In the large $N_c$ limit, the light
and singly heavy baryons are viewed respectively as $N_c$ and $N_c-1$
valence quarks bound by the pion mean fields created
self-consistently, since a heavy quark can be regarded as a static
color source in the limit of the infinitely heavy quark mass.
The transition magnetic moments of the baryon sextet are 
determined entirely by using the parameters fixed in the light-baryon
sector without any additional parameters introduced. Assuming that the
transition $E2$ moments are small, we are able to compute the
radiative decay rates of the baryon sextet.  The numerical
results are discussed, being compared with those from other
approaches.   
\end{abstract}
\maketitle
\section{Introduction}
In the large $N_c$ (the number of colors) limit, an ordinary baryon
can be viewed as a state of $N_c$ valence quarks bound by 
the meson mean fields, which are self-consistently created by the
presence of the $N_c$ valence
quarks~\cite{Witten:1979kh,Witten:1983tx}. The chiral quark-soliton
model ($\chi$QSM)~\cite{Diakonov:1987ty} realizes effectively this
picture (for a review, we refer to
Refs.~\cite{Christov:1995vm,Diakonov:1997sj}).  On an equal footing,  
a singly heavy baryon, which consists of one heavy quark and two light
quarks, can be regarded as a state of $N_c-1$ valence quarks bound by
the pion mean fields, if the heavy-quark mass is taken to be infinity
($m_Q\to \infty$)~\cite{Diakonov:2010tf, Yang:2016qdz} (see also a
recent review~\cite{Kim:2018cxv}). The masses of
the lowest-lying singly heavy baryons in both the charmed and bottom
flavor sectors were successfully described within the framework of
this mean-field approach, and in particular the mass of the
$\Omega_{b}$ was predicted~\cite{Yang:2016qdz, Kim:2018xlc} with all
parameters fixed in the light baryon sector. More recently, 
the five narrow $\Omega_c$ resonances newly observed by
the LHCb Collaboration~\cite{Aaij:2017nav} were studied 
within the same framework and two of them were 
advocated as exotic pentaquark baryons belonging to the
anti-decapentaplet ($\overline{\bm{15}}$)~\cite{Kim:2017jpx}. 
This classification of the narrow $\Omega_c$ resonances as the members
of the baryon anti-decapentaplet has been further supported by
computing strong decays of the newly found two narrow
$\Omega_{c}$s~\cite{Kim:2017khv}. The magnetic moments of singly 
heavy baryons have been also investigated without any new additional 
parameters~\cite{Yang:2018uoj}.  Very recently, the electromagnetic
form factors of the singly heavy baryons have been also computed
within the self-consistent $\chi$QSM~\cite{Kim:2018nqf}. 

While there is no experimental information on radiative decays of
heavy baryons, the CLEO Collaboration identified the two resonances of
the singly heavy baryons, $\Xi_{c}^{+\prime}$ and $\Xi_c^{0\prime}$,
using their radiative decays $\Xi_{c}^{+\prime}\to \Xi_{c}^{+}\gamma$
and $\Xi_c^{0\prime}\to \Xi_{c}^{0}\gamma$, since the masses of the 
$\Xi_c^{\prime}$ isodoublet lie below threshold for their strong
decays~\cite{Jessop:1998wt}. Theoretically, radiative 
decays of singly heavy baryons have been studied within various
different approaches: The nonrelativistic quark model
(NRQM)~\cite{Johnson:1976mv, Choudhury:1976dp, Lichtenberg:1976fi,
  Franklin:1981rc}, the bag models~\cite{Bose:1980vy,
  Bernotas:2013eia}, potential models~\cite{Jena:1986xs,
  Majethiya:2009vx}, QCD sum rules~\cite{Aliev:2001xr, Aliev:2009jt,
  Aliev:2014bma, Aliev:2016xvq}, heavy-quark effective
theories~\cite{Cheng:1992xi, Zhu:1998ih}, chiral perturbation
theories~\cite{Banuls:1999br, Jiang:2015xqa, 
  Wang:2018cre}, relativistic quark models~\cite{Ivanov:1998wj}, the
heavy quark symmetry with a diquark picture~\cite{Tawfiq:1999cf}, the
chiral constituent quark model~\cite{Sharma:2010vv, Wang:2017kfr}, and
lattice QCD~\cite{Bahtiyar:2015sga, Bahtiyar:2016dom}.
In the present work, we want to investigate the transition magnetic
moments and radiative decays of singly heavy baryons belonging to the
baryon sextet with both spin 1/2 and 3/2. The main virtue of the
present approach lies in the fact that all the dynamical parameters,
which are required to compute the transition magnetic moments of the
singly heavy baryons, have already been fixed in the light-baryon
sector~\cite{Yang:2015era, Yang:2018uoj}. Thus, we can predict the
transition magnetic moments and the radiative decay rates of the
lowest-lying singly-heavy baryons in a robust manner. 

The present work is organized as follows: In Section
II, we explain a general formalism of the $\chi$QSM to compute the
transition magnetic moments of the heavy baryons. 
In Section III, we explicitly calculate the transition magnetic
moments within the present framework. In Section IV, we present the
numerical results of the transition magnetic moments and radiative
decays of the heavy baryons. The last Section is devoted to the
summary and conclusions of the present work.

\section{General formalism}
We first introduce the electromagnetic (EM) current that consists of both
the light and heavy quark currents
\begin{align}
  J_{\mu}(x) = \bar{\psi}(x)\gamma_{\mu}\hat{\mathcal{Q}} \psi(x)
  + e_{Q}\bar{Q}\gamma_{\mu}Q, 
\label{eq:LHcurrent}
\end{align}
where $\hat{\mathcal{Q}}$ stands for the charge operator of the light
quarks, defined as 
\begin{align}
\hat{\mathcal{Q}}=
\begin{pmatrix}\frac{2}{3} 
& 0 
& 0\\
  0 
& -\frac{1}{3} 
& 0\\
  0 
& 0 
& -\frac{1}{3}
\end{pmatrix}
=\frac{1}{2}\left(\lambda_{3}+\frac{1}{\sqrt{3}}\lambda_{8}\right).
\label{eq:chargeOp}
\end{align}
Here, $\lambda_{3}$ and $\lambda_{8}$ represent the well-known flavor
SU(3) Gell-Mann matrices. The second part of Eq.~(\ref{eq:LHcurrent})
is the heavy-quark  EM current with heavy-quark charge $e_Q$
($e_{c}=2/3$ for the charm quark or as $e_{b}=-1/3$ for the bottom
quark). However, the heavy-quark EM current is not involved in the
infinitely heavy-quark mass limit $m_Q\to\infty$, since the magnetic
moment of a heavy quark is proportional to the inverse of the
corresponding mass,  i.e. $\bm{\mu}\sim(e_{Q}/m_{Q})\bm{\sigma}$.
We expect that the corrections from the next-to-leading order in the
$1/m_Q$ expansion should be rather small in comparison with the
light-quark contributions to the transition magnetic moments. 
So, we consider only the first term of Eq.~\eqref{eq:LHcurrent}
when we calculate the transition magnetic moments of heavy baryons.
Since we employ the limit of $m_Q\to \infty$, we obtain the
same results for both the charmed and bottom baryons, which is the
consequence of heavy flavor symmetry. 

Starting from the transition matrix element of the EM current
sandwiched between the heavy-baryon states, we are able to derive the
collective operator for the magnetic moments within the framework of
the $\chi$QSM. In fact, they have been already derived in previous 
works~\cite{Kim:1995mr,Kim:1995ha,Kim:1997ip, Kim:2005gz,
  Yang:2015era}. Considering the rotational $1/N_{c}$ and linear
strange current-quark mass $m_{\mathrm{s}}$ corrections, we obtain the
collective operator for the magnetic moments as 
\begin{align}
\hat{\mu}=\hat{\mu}^{(0)}+\hat{\mu}^{(1)},
\label{eq:MagMomOp}
\end{align}
where $\hat{\mu}^{(0)}$ and $\hat{\mu}^{(1)}$ denote the leading
and rotational $1/N_{c}$ contributions, and the linear $m_{\mathrm{s}}$
corrections respectively, which are expressed as  
\begin{align}
\hat{\mu}^{(0)} & =
\;\; w_{1}D_{\mathcal{Q}3}^{(8)}
\;+\;w_{2}d_{pq3}D_{\mathcal{Q}p}^{(8)}\cdot\hat{J}_{q}
\;+\;\frac{w_{3}}{\sqrt{3}}D_{\mathcal{Q}8}^{(8)}\hat{J}_{3},
\\
\hat{\mu}^{(1)} & =
\;\;\frac{w_{4}}{\sqrt{3}}d_{pq3}D_{\mathcal{Q}p}^{(8)}D_{8q}^{(8)}
+ w_{5}
\left(D_{\mathcal{Q}3}^{(8)}D_{88}^{(8)}+D_{\mathcal{Q}8}^{(8)}D_{83}^{(8)}\right) 
\;+\;  w_{6}
\left(D_{\mathcal{Q}3}^{(8)}D_{88}^{(8)}-D_{\mathcal{Q}8}^{(8)}D_{83}^{(8)}\right).  
\label{eq:magop}
\end{align}
The indices of symmetric tensor $d_{pq3}$ run over $p=4,\cdots,\,7$.
$\hat{J_{3}}$ and $\hat{J}_{p}$ correspond respectively to the third
and the $p$th components of the spin operator acting on the
soliton. $D_{ab}^{(\nu)}(R)$ are the SU(3) Wigner matrices in
the representation $\nu$, which arise from the quantization of the
soliton. $D_{\mathcal{Q}3}^{(8)}$ is defined by the combination of the
SU(3) Wigner $D$ functions  
\begin{align}
D_{\mathcal{Q}3}^{(8)}
=\frac{1}{2}\left(D_{33}^{(8)}+\frac{1}{\sqrt{3}}D_{83}^{(8)}\right),
\end{align}
which comes from the SU(3) rotation of the EM octet current. The
coefficients $w_{i}$ in Eq.~\eqref{eq:magop} contain  
a specific dynamics of the chiral soliton and are independent of baryons
involved. $w_{1,2,3}$ are the SU(3) symmetric parts that are of order
$\mathcal{O}(m_s^0)$. Though $w_1$ includes the $m_s$ correction,
it is not explicitly involved in the breaking of flavor SU(3)
symmetry. So, we will consider $w_{1}$ as the SU(3) symmetric part,
when the transition magnetic moments are derived. 
On the other hand, $w_{4,5,6}$ are of the leading order in the $m_s$
expansion, i.e. of order $\mathcal{O}(m_s)$. The dynamical
coefficients $w_{i}$ can be explicitly computed within a specific chiral 
solitonic model such as the $\chi$QSM~\cite{Kim:1995mr,Kim:1995ha,
  Yang:2018uoj}. 

The structure of the collective operator $\hat{\mu}$ in
Eq.~(\ref{eq:magop}) is in fact very general and is considered as a 
\emph{model-independent} one. It is deeply related to hedgehog 
symmetry and the embedding of the SU(2) soliton into
SU(3)~\cite{Witten:1983tx}, which provides a minimal way of combining
the ordinary space with the internal flavor space. Because of this
embedding, the symmetry we have is
$\mathrm{SU(2)}_{T}\times\mathrm{U(1)}_{Y}$. Therefore, the collective  
operator for the magnetic moment is expressed in terms of 
the $\mathrm{SU(2)}_{T}\times\mathrm{U(1)}_{Y}$ invariant tensors
\begin{align}
d_{abc}=\frac{1}{4}\mathrm{tr}(\lambda_{a}\{\lambda_{b},\,\lambda_{c}\}),
\;\;\;
S_{ab3}=\sqrt{\frac{1}{3}}(\delta_{a3}\delta_{b8}+\delta_{b3}\delta_{a8}),
\;\;\;
F_{ab3}=\sqrt{\frac{1}{3}}(\delta_{a3}\delta_{b8}-\delta_{b3}\delta_{a8}), 
\end{align}
which yield the general expression for the collective operator given in
Eq.~\eqref{eq:magop}. Hence, instead of computing $w_i$ in a
specific model, we can determine $w_{i}$ by the experimental
data on the magnetic moments of the baryon octet as done in
Refs.~\cite{Kim:1997ip,Yang:2015era,Kim:2005gz}.

To derive the transition magnetic moments of the heavy baryons, we
need to compute the matrix elements of the collective operator
$\hat{\mu}$ in Eq.~\eqref{eq:MagMomOp} between heavy baryon
states. Moreover, the presence of $w_{4,5,6}$ comes from the
perturbative treatment of $m_{s}$, so that
the collective wavefunctions for the soliton consisting of the
light-quark pair are not pure states but mixed ones with higher
representations. In order to derive the collective wavefunctions, we
need to diagonalize the collective Hamiltonian for flavor SU(3)
symmetry breaking, expressed as 
\begin{align}
H_{\mathrm{sb}}
=\alpha
  D_{88}^{(8)}+\beta\hat{Y}+\frac{\gamma}{\sqrt{3}}\sum_{i=1}^{3}
  D_{8i}^{(8)}\hat{J}_{i}.  
\label{eq:Hamiltonian}
\end{align}
where $\alpha$, $\beta$, and $\gamma$ are the dynamical coefficients
for the lowest-lying singly heavy baryons, of which the explicit
expressions can be found in Refs.~~\cite{Yang:2016qdz, Kim:2018xlc}. 
Diagonalizing Eq.~\eqref{eq:Hamiltonian}, we obtain the wavefunctions
for the baryon anti-triplet ($J=0$) and the sextet ($J=1$)
respectively~\cite{Kim:2018xlc} as 
\begin{align}
 & |B_{\overline{\bm{3}}_{0}}\rangle
=|\overline{\bm{3}}_{0},B\rangle+p_{\overline{15}}^{B}|
   \overline{\bm{15}}_{0},B\rangle,  
\cr
 & |B_{\bm{6}_{1}}\rangle
=|{\bm{6}}_{1},B\rangle+q_{\overline{15}}^{B}
|{\overline{\bm{15}}}_{1},B\rangle+q_{\overline{24}}^{B}|
{{\overline{\bm{24}}}_{1}},B\rangle,
\label{eq:mixedWF1}
\end{align}
with the mixing coefficients 
\begin{align}
p_{\overline{15}}^{B}\;\;=\;\;p_{\overline{15}}
\left[
\begin{array}{c}
-\sqrt{15}/10\\
-3\sqrt{5}/20
\end{array}
\right], 
\hspace{3em}
q_{\overline{15}}^{B}\;\;=\;\;q_{\overline{15}}
\left[
\begin{array}{c}
\sqrt{5}/5\\
\sqrt{30}/20\\
0
\end{array}
\right], 
\hspace{3em}
q_{\overline{24}}^{B}\;\;=\;\;q_{\overline{24}}
\left[
\begin{array}{c}
-\sqrt{10}/10\\
-\sqrt{15}/10\\
-\sqrt{15}/10
\end{array}
\right],
\label{eq:pqmix-1}
\end{align}
respectively, in the basis $\left[\Lambda_{Q},\;\Xi_{Q}\right]$ for
the anti-triplet and
$\left[\Sigma_{Q}\left(\Sigma_{Q}^{\ast}\right),
\;\Xi_{Q}^{\prime}\left(\Xi_{Q}^{\ast}\right),
\;\Omega_{Q}\left(\Omega_{Q}^{\ast}\right)\right]$
for the sextets. The expressions for the parameters
$p_{\overline{15}}$, $q_{\overline{15}}$, and $q_{\overline{24}}$ are
also found in Refs.~\cite{Yang:2018uoj, Kim:2018xlc} and 
the corresponding numerical values are given as 
\begin{align}
p_{\overline{15}} \;=\; -0.104 \pm 0.003, \;\;\;
q_{\overline{15}} \;=\;  0.238 \pm 0.005, \;\;\;
q_{\overline{24}} \;=\; -0.049 \pm 0.002.
\label{eq:pqmix_values}
\end{align}
Note that the mixing coefficients are proportional to $m_s$
linearly. So, the effects of flavor SU(3) symmetry breaking arise also
from the collective wavefunctions for the heavy baryons in addition to
the collective operator for the magnetic moments, which contains
$w_{4,5,6}$. 

The complete wavefunction for a heavy baryon can be constructed by
coupling the soliton wavefunction to the heavy quark such that the
heavy baryon becomes a color singlet, which are expressed as 
\begin{align}
|B_\mu; (J',J_3')(T,T_3) \rangle 
=\sum_{J_{3},\,J_{Q3}}C_{\,J,J_{3}\,J_{Q}\,J_{Q3}}^{J'\,J_{3}'}
  \;\mathbf{\chi}_{J_{Q3}}\; 
|B_{\mu}; (J,J_3)(T,T_3)\rangle
\label{eq:HeavyWF-1}
\end{align}
where $\chi_{J_{Q3}}$ stand for the Pauli spinors and 
$C_{\,J,J_{3}\,J_{Q}\,J_{Q3}}^{J'\,J_{3}'}$ denote the Clebsch-Gordan 
coefficients. $J'$ and $J_3'$ represent the spin and its third
component of the heavy baryons whereas $T$ and $T_3$ are the
corresponding isospin and its third component. The wavefunctions
$|B_{\mu}; (J,J_3)(T,T_3)\rangle$ with representation $\mu$ are given in
Eq.~\eqref{eq:mixedWF1}. Calculating the matrix elements of
$\hat{\mu}$ becomes just the simple $D$-function algebra. 
Thus, we can easily obtain the transition magnetic moments of the 
heavy baryons.

In Ref.~\cite{Yang:2018uoj}, we discussed in detail how $w_1$, $w_4$,
$w_5$, and $w_6$ are modified for the singly heavy baryons that
consist of $N_c-1$ light valence quarks. Accordingly, the first
coefficient $w_1$ was revised as 
\begin{align}
\widetilde{w}_{1}
=\left[\frac{N_{c}-1}{N_{c}}(w_{1}+w_{3})-w_{3}\right]\sigma,
\label{eq:w1tilde}
\end{align}
where $\sigma$ is introduced to compensate the deviation arising from
the nonrelativistic limit used in the course of deriving 
$\widetilde{w}_1$. The numerical value of $\sigma$ was already
determined in Ref.~~\cite{Kim:2017khv}: $\sigma\sim0.85$. 
The other two coefficients $w_{2}$ and $w_{3}$ are
kept to be intact in the course of reducing the number of valence
quarks from $N_c$ to $N_c-1$ whereas $w_{4,5,6}$ need to be
modified as
\begin{align}
\overline{w}_{i}
=\frac{(N_{c}-1)}{N_{c}}w_{i},\;\;\;i=4,\,5,\,6.
\label{eq:tildeW}
\end{align}
Using the numerical values of $w_i$ determined in
Ref.~\cite{Yang:2015era} and Eqs.~\eqref{eq:w1tilde} and
\eqref{eq:tildeW}, we obtain the following values~\cite{Yang:2018uoj}  
\begin{align}
\widetilde{w}_{1} & -=10.08\pm0.24,\cr
w_{2} & =4.15\pm0.93,\cr
w_{3} & =8.54\pm0.86,\cr
\overline{w}_{4} & =-2.53\pm0.14,\cr
\overline{w}_{5} & =-3.29\pm0.57,\cr
\overline{w}_{6} & =-1.34\pm0.56.
\label{eq:numW}
\end{align}
Thus, once we know the numerical values of these dynamical parameters,
we can straightforwardly compute the transition magnetic moments of
singly heavy baryons, which will be shown in the next Section.  

\section{Transition magnetic moments and partial decay
  widths of radiative decays}
The matrix elements of the EM current between the baryon sextet with
spin $1/2$  to the baryon antitriplet can be parametrized by the
transition EM form factors $F_{i}(q^{2})$ as follows: 
\begin{align}\
\langle B_{1/2} ^\prime (p^{\prime})|J_{\mu}(0)|B_{1/2}(p)\rangle
=\overline{u} _{B_{1/2} ^\prime} ({\bm{p}}^{\prime},\lambda^{\prime})\left[F_{1}(q^{2})
+i\sigma_{\mu\nu}\frac{q^{\nu}}{ M_{B_{1/2} ^\prime}+M_{B_{1/2}}} F_{2}\right]
u_{B_{1/2}}({\bm{p}},\lambda),
\label{eq:em1}
\end{align}
where $q^{2}$ is the square of the four-momentum transfer, and
$M_{B_{1/2} ^\prime}$ and $M_{B_{1/2}}$ denote respectively the masses 
of the baryon antitriplet and sextet with spin $1/2$.
The $u_{B(B')_{1/2}}$ is the Dirac spinor corresponding to
the baryon $B(B')$. Similarly, the transition EM form factors from the 
baryon sextet with spin $3/2$ to the sextet or antitriplet with spin
$1/2$, i.e.  ${\bf 6}_{3/2} \rightarrow {\bm{\overline{3}}}_{1/2}$ or 
  $\bm{6}_{3/2} \rightarrow \bm{6}_{1/2}$, are defined
  as~\cite{Jones:1972ky}   
\begin{align}
\langle B_{1/2}(p^{\prime})|J_{\mu}|B_{3/2}(p)\rangle 
 =  
i\sqrt{\frac{2}{3}}\bar{u}_{B_{1/2}}({\bm{p}}^{\prime},\lambda^{\prime})
\left[G_{M}^{*}(q^{2})\mathcal{K}_{\beta\mu}^{M}
+G_{E}^{*}(q^{2})\mathcal{K}_{\beta\mu}^{E}  
+G_{C}^{*}(q^{2})\mathcal{K}_{\beta\mu}^{C}\right]
u_{B_{3/2}}^{\beta}({\bm{p}},\Lambda),
\label{eq:em2}
\end{align}
where $G_{M}^{*}$, $G_{E}^{*}$, and $G_{C}^{*}$ stand respectively for 
the transition magnetic dipole, electric quadrupole, and Coulomb form
factors. The $\mathcal{\ K}^{M,E,C}$ are the covariant tensors defined
in Ref.~\cite{Jones:1972ky}: 
\begin{align}
\mathcal{K}_{\beta\mu}^{M} 
& =  -i\frac{3(M_{3/2}+M_{1/2})}{2M_{1/2}[(M_{3/2}+M_{1/2})^{2}-q^{2}]}
\epsilon_{\beta\mu\lambda\sigma}P^{\lambda}q^{\sigma},\cr
{\cal K}_{\beta\mu}^{E} 
& = 
-\mathcal{K}_{\beta\mu}^{M}-\frac{6(M_{3/2}+M_{1/2})}{M_{1/2}\Delta(q^{2})}
\epsilon_{\beta\sigma\lambda\rho}P^{\lambda}q^{\rho}
\epsilon_{\mu\kappa\delta}^{\sigma}P^{\kappa}q^{\delta}\gamma^{5},\cr
{\cal K}_{\beta\mu}^{C} 
& = 
-i\frac{3(M_{3/2}+M_{1/2})}{M_{1/2}\Delta(q^{2})}q_{\beta}(q^{2}P_{\mu}
-q\cdot Pq_{\mu})\gamma^{5}
\end{align}
with 
\begin{align}
\Delta(q^{2})=[(M_{3/2}+M_{1/2})^{2}-q^{2}][(M_{3/2}-M_{1/2})^{2}-q^{2}].
\end{align}
$u_{B_{3/2}}^{\beta}$ represents the Rarita-Schwinger spinor for the
baryon sextet with spin 3/2. At $q^{2}=0$, note that the transition
magnetic dipole form factors $F_{2}(q^{2})$ and $G_{M}^{*}(q^{2})$ are
the same as the transition magnetic moments. Since the magnetic dipole 
transitions ($M1$) are experimentally dominant over the electric
quadrupole transitions ($E2$) in hyperon radiative decays, one can
neglect the contribution from the $E2$ transitions to the radiative
decays of heavy baryons. This is a plausible approximation, since the
the size of the $E2/M1$ ratio for the $\Delta$ isobar is yielded in
the range of $(1-3)\,\%$ experimentally~\cite{Blanpied:2001ae,
  Frolov:1998pw, Joo:2001tw, Mertz:1999hp, Beck:1999ge,
  Pospischil:2000ad}. Though there is no experimental data on the
$E2/M1$ ratio for the singly heavy baryons, we can safely assume that
the size of the $E2/M1$ ratio is very small. It indicates that the
effects of the $E2$ are expected to be negligible on the radiative
decay widths of the singly heavy baryons, because even the squared
values of the $E2$ contribute to them. In fact, a recent lattice study
has predicted the magnitude of the $E2$ form factor of $\Omega_c^*$ to
be much smaller than those of the $M1$ form
factors~\cite{Bahtiyar:2015sga}: $M1(\Omega_c^*)=-0.657$ and
$E2(\Omega_c^*)=-0.012$. A very recent calculation of the
electromagnetic transition form factors of the singly heavy baryons
within the self-consistent SU(3) $\chi$QSM has come to a similar
conclusion~\cite{kimkim}.  

So, we can express the radiative decay rates in terms of the
transition magnetic moments. Using Eqs.~(\ref{eq:em1},\ref{eq:em2})
and neglecting the $E2$ transitions, we obtain the radiative decay
rates for $B_{1/2} \rightarrow B_{1/2} ^\prime$ and  for $B_{3/2}
\rightarrow B_{1/2} ^\prime$, respectively:  
\begin{align}
\Gamma(B_{1/2}\to B_{1/2}^{\prime}\gamma) 
& =  4\alpha_{\mathrm{EM}}\frac{E_{\gamma}^{3}}{(M_{B_{1/2}^{\prime}}
+M_{B_{1/2}})^{2}}\left(\frac{\mu_{B_{1/2}^{\prime}B_{1/2}}}{\mu_{N}}\right)^{2},
\label{eq:parwidth1}\\
\Gamma(B_{3/2}\to B_{1/2}^{\prime}\gamma) 
& = 
\frac{\alpha_{\mathrm{EM}}}{2}\frac{E_{\gamma}^{3}}{M_{B_{1/2}^{\prime}}^{2}}
\left(\frac{\mu_{B_{1/2}^{\prime}B_{3/2}}}{\mu_{N}}\right)^{2},
\label{eq:parwidth2}
\end{align}
respectively, where $M_{B_{J}}$ is the mass of baryon $B$ with spin
$J$ and $\mu_{B_{J^{\prime}}^{\prime}B_{J}}$ is the transition magnetic
moments for the radiative decay $B_{J}\rightarrow
B_{J^{\prime}}^{\prime}\gamma$.  
In Eq.(\ref{eq:parwidth1},\ref{eq:parwidth2}) $\alpha_{\mathrm{EM}}$
denotes the fine structure constant and $E_{\gamma}$ the energy
of the produced photon: 
\begin{align}
E_{\gamma}=\frac{M_{B_{J}}^{2}-M_{B_{J^{\prime}}^{\prime}}^{2}}{2M_{B_{J}}}.
\label{eq:E_gamma}
\end{align}

Sandwiching the collective operator for the magnetic moment given in
Eq.~\eqref{eq:MagMomOp}  between the heavy baryon states expressed by
the collective wavefunctions \eqref{eq:HeavyWF-1}, we finally arrive
at the expressions of the transition magnetic moments for the baryon
from the baryon sextet with spin $1/2$ to the antitriplet, ${\bf
  6}_{1/2} \rightarrow {\overline{\mathbf{3}}}_{1/2}$,  as 
\begin{align}
\mu^{\left(0\right)}\left[\Sigma_{c}^{+}\rightarrow\Lambda_{c}^{+}\right] 
& =  
-\frac{1}{4\sqrt{6}}\left(\widetilde{w}_{1}-\frac{1}{2}w_{2}\right),
\cr
\mu^{\left(\mathrm{op}\right)}\left[\Sigma_{c}^{+}\rightarrow\Lambda_{c}^{+}\right] 
& = 
-\frac{1}{20\sqrt{6}}\left(\overline{w}_{4}+2\overline{w}_{5}+\overline{w}_{6}\right),
\cr
\mu^{\left(\mathrm{wf}\right)}\left[\Sigma_{c}^{+}\rightarrow\Lambda_{c}^{+}\right] 
& =  
\frac{1}{120\sqrt{2}}\left(\widetilde{w}_{1}+\frac{3}{2}w_{2}\right)p_{\overline{15}}
\;-\;\frac{1}{20\sqrt{3}}\left(\widetilde{w}_{1}+\frac{1}{2}w_{2}\right)q_{\overline{15}},
\cr 
\mu^{\left(0\right)}\left[\Xi_{c}^{\prime}\rightarrow\Xi_{c}\right] 
& =  \frac{1}{4\sqrt{6}}\mathcal{Q}\left(\widetilde{w}_{1}-\frac{1}{2}w_{2}\right),
\cr
\mu^{\left(\mathrm{op}\right)}\left[\Xi_{c}^{\prime}\rightarrow\Xi_{c}\right] 
& = 
\frac{1}{20\sqrt{6}}\left[\left(\mathcal{Q}-1\right)\overline{w}_{4}
+2\left(\mathcal{Q}-1\right)\overline{w}_{5}
-\left(2\mathcal{Q}-1\right)\overline{w}_{6}\right],
\cr
\mu^{\left(\mathrm{wf}\right)}\left[\Xi_{c}^{\prime}\rightarrow\Xi_{c}\right] 
& =  -\frac{1}{240\sqrt{2}}(5\mathcal{Q}-4)\left(\widetilde{w}_{1}+\frac{3}{2}w_{2}
\right)p_{\overline{15}}
\;+\;\frac{1}{80\sqrt{3}}(\mathcal{Q}-2)\left(\widetilde{w}_{1} +
 \frac{1}{2}w_{2}\right)q_{\overline{15}}, 
\label{eq:61_3bar_2}
\end{align}
where $\mathcal{Q}$ designates the electric charges of the
corresponding heavy baryons. $\mu^{(0)}[B_c\to B_c' ]$ denotes the
contribution from the leading order. $\mu^{(\mathrm{op})}$ stands for
the linear $m_s$ terms from the collective operator~~\eqref{eq:magop}
and $\mu^{(\mathrm{wf})}$ comes from the collective
wavefunctions~\eqref{eq:mixedWF1}.  

Similarly, we can derive the expressions of the transition magnetic
moments for the $\mathbf{6}_{3/2} \rightarrow
{\overline{\mathbf{3}}}_{1/2} \gamma$ decays as    
\begin{align}
\mu^{\left(0\right)}\left[\Sigma_{c}^{\ast+}\rightarrow\Lambda_{c}^{+}\right] 
& =  \frac{1}{4\sqrt{3}}\left(\widetilde{w}_{1}-\frac{1}{2}w_{2}\right),
\cr
\mu^{\left(\mathrm{op}\right)}\left[\Sigma_{c}^{\ast+}
\rightarrow\Lambda_{c}^{+}\right]  
& = 
\frac{1}{20\sqrt{3}}\left(\overline{w}_{4}+2\overline{w}_{5}
+\overline{w}_{6}\right),
\cr
\mu^{\left(\mathrm{wf}\right)}\left[\Sigma_{c}^{\ast+}\rightarrow
\Lambda_{c}^{+}\right] 
& = 
-\frac{1}{120}\left(\widetilde{w}_{1}+\frac{3}{2}w_{2}\right)
p_{\overline{15}}
\;+\;\frac{1}{10\sqrt{6}}\left(\widetilde{w}_{1}
+\frac{1}{2}w_{2}\right)q_{\overline{15}},
\cr 
\mu^{\left(0\right)}\left[\Xi_{c}^{\ast}\rightarrow\Xi_{c}\right] 
& =  -\frac{1}{4\sqrt{3}}\mathcal{Q}\left(\widetilde{w}_{1}
-\frac{1}{2}w_{2}\right),\cr
\mu^{\left(\mathrm{op}\right)}\left[\Xi_{c}^{\ast}\rightarrow\Xi_{c}\right] 
& = 
-\frac{1}{20\sqrt{3}}\left[(\mathcal{Q}-1)\overline{w}_{4}
+2(\mathcal{Q}-1)\overline{w}_{5}-(2\mathcal{Q}-1)\overline{w}_{6}\right],
\cr
\mu^{\left(\mathrm{wf}\right)}\left[\Xi_{c}^{\ast}\rightarrow\Xi_{c}\right] 
& = 
\frac{1}{240}(5\mathcal{Q}-4)\left(\widetilde{w}_{1}
+\frac{3}{2}w_{2}\right)p_{\overline{15}}
\;-\;\frac{1}{40\sqrt{6}}(\mathcal{Q}-2)
\left(\widetilde{w}_{1}+\frac{w_{2}}{2}\right)q_{\overline{15}},
\label{eq:63_3bar_2}
\end{align}
and for ${\bf 6}_{3/2} \rightarrow {\bf 6}_{1/2} \gamma$ as 
\begin{align}
\mu^{\left(0\right)}\left[\Sigma_{c}^{\ast}\rightarrow\Sigma_{c}\right] 
& = -\frac{1}{30\sqrt{2}}\left(3\mathcal{Q}-2\right)
\left(\widetilde{w}_{1}-\frac{1}{2}w_{2}-\frac{1}{3}w_{3}\right),
\cr
\mu^{\left(\mathrm{op}\right)}\left[\Sigma_{c}^{\ast}
\rightarrow\Sigma_{c}\right] 
& = 
-\frac{1}{270\sqrt{2}}\left[(5\mathcal{Q}-7)\overline{w}_{4}
+3(4\mathcal{Q}-5)\overline{w}_{5}\right],
\cr
\mu^{\left(\mathrm{wf}\right)}\left[\Sigma_{c}^{\ast}\rightarrow
\Sigma_{c}\right] 
& = -\frac{1}{45}(\mathcal{Q}-2)\left(\widetilde{w}_{1}
+\frac{1}{2}w_{2}+w_{3}\right)
q_{\overline{15}}  \;+\;\frac{1}{180\sqrt{5}}(\mathcal{Q}+1)
\left(\widetilde{w}_{1}
+2w_{2}-2w_{3}\right)q_{\overline{24}},\cr 
\mu^{\left(0\right)}\left[\Xi_{c}^{\ast}\rightarrow
\Xi_{c}^{\prime}\right] 
& = 
-\frac{1}{30\sqrt{2}}\left(3\mathcal{Q}-2\right)
\left(\widetilde{w}_{1}-\frac{1}{2}w_{2}-\frac{1}{3}w_{3}\right),
\cr
\mu^{\left(\mathrm{op}\right)}\left[\Xi_{c}^{\ast}
\rightarrow\Xi_{c}^{\prime}\right] 
& = 
-\frac{1}{270\sqrt{2}}\left[(7\mathcal{Q}-2)\overline{w}_{4}
+(6\mathcal{Q}-3)\overline{w}_{5}\right],
\cr
\mu^{\left(\mathrm{wf}\right)}\left[\Xi_{c}^{\ast}
\rightarrow\Xi_{c}^{\prime}\right] 
& = 
-\frac{1}{180}(5\mathcal{Q}-4)\left(\widetilde{w}_{1}
+\frac{1}{2}w_{2}+w_{3}\right)q_{\overline{15}}
\;+\;\frac{1}{90\sqrt{5}}(\mathcal{Q}+1)
\left(\widetilde{w}_{1}+2w_{2}-2w_{3}\right)q_{\overline{24}},\cr
\mu^{\left(0\right)}\left[\Omega_{c}^{\ast}\rightarrow\Omega_{c}\right] 
& = 
\frac{1}{15\sqrt{2}}\left(\widetilde{w}_{1}-\frac{1}{2}w_{2}
-\frac{1}{3}w_{3}\right),
\cr
\mu^{\left(\mathrm{op}\right)}\left[\Omega_{c}^{\ast}
\rightarrow\Omega_{c}\right] 
& =  
-\frac{1}{90\sqrt{2}}\left(\overline{w}_{4}
+3\overline{w}_{5}\right),\cr
\mu^{\left(\mathrm{wf}\right)}\left[\Omega_{c}^{\ast}
\rightarrow\Omega_{c}\right] 
& = 
\frac{1}{60\sqrt{5}}\left(\widetilde{w}_{1}+2w_{2}-2w_{3}
\right)q_{\overline{24}}.
\label{eq:63_61_c}
\end{align}

As for the transition magnetic moments of the baryon sextet to the
antitriplet, we find the following general relations
\begin{align}
\frac{\mu\left[{\bf 6}_{3/2} \rightarrow 
  \overline{\mathbf{3}}_{1/2}\right]}{\mu\left[{\bf 6}_{1/2} \rightarrow 
  \overline{\mathbf{3}}_{1/2}\right]}  
\, =\,  
\frac{\mu\left[\Sigma_{c}^{\ast+}
\rightarrow\Lambda_{c}^{+}\right]}{\mu\left[\Sigma_{c}^{+}
\rightarrow\Lambda_{c}^{+}\right]} 
\;\; = \;\; 
\frac{\mu\left[\Xi_{c}^{\ast}
\rightarrow\Xi_{c}\right]}{\mu\left[\Xi_{c}^{\prime}
\rightarrow\Xi_{c}\right]}
\;\;=\;\;-\sqrt{2}.
\label{eq:rel1}
\end{align}
These relations are satisfied even with the effects of flavor SU(3)
symmetry breaking considered.  
Note that the NRQM~\cite{Johnson:1976mv, Choudhury:1976dp,
  Lichtenberg:1976fi} also satisfy Eq.~\eqref{eq:rel1}.
In the chiral limit, the $U$-spin symmetry yields various relations: 
\begin{align}
\mu\left[\Sigma_{c}^{\ast++}\rightarrow\Sigma_{c}^{++}\right] 
& =  
4\,\mu\left[\Sigma_{c}^{\ast+}\rightarrow\Sigma_{c}^{+}\right]
\;\;=\;\;
-2\,\mu\left[\Sigma_{c}^{\ast0}\rightarrow\Sigma_{c}^{0}\right]
\cr
=\;\;
4\,\mu\left[\Xi_{c}^{\ast+}\rightarrow\Xi_{c}^{\prime+}\right] 
& =  
-2\,\mu\left[\Xi_{c}^{\ast0}\rightarrow\Xi_{c}^{\prime0}\right]
\;\;=\;\;
-2\,\mu\left[\Omega_{c}^{\ast0}\rightarrow\Omega_{c}^{0}\right],
\cr
\mu\left[\Sigma_{b}^{\ast+}\rightarrow\Sigma_{b}^{+}\right] 
& =  4\,\mu\left[\Sigma_{b}^{\ast0}\rightarrow\Sigma_{b}^{0}\right]
\;\;=\;\;
-2\,\mu\left[\Sigma_{b}^{\ast-}\rightarrow\Sigma_{b}^{-}\right]
\cr
=\;\;
4\,\mu\left[\Xi_{b}^{\ast0}\rightarrow\Xi_{b}^{\prime0}\right] 
& =  
-2\,\mu\left[\Xi_{b}^{\ast-}\rightarrow\Xi_{b}^{\prime-}\right]
\;\;=\;\;
-2\,\mu\left[\Omega_{b}^{\ast-}\rightarrow\Omega_{b}^{-}\right].
\label{eq:sumR2}  
\end{align}
The similar relations are obtained in Refs.~\cite{Banuls:1999br,
  Wang:2018cre}, in which recombinations of Eq.~\eqref{eq:sumR2} are
found.  

\section{Results and discussion}
We now present the numerical results of the transition magnetic
moments for the lowest-lying singly heavy baryons.
As mentioned previously, all the dynamical parameters were already
fixed by using the experimental data on the magnetic moments of the
baryon octet. We want to emphasize that the present approach describes
successfully the magnetic moments and transition magnetic moments for
the baryon decuplet with those parameters~\cite{Kim:2017khv} only.  
So, the present results for heavy baryons are obtained
without any additional free parameters.
In Table~\ref{tab:1}, we list the numerical results of the transition
magnetic moments for the charmed baryons. The third column lists those
in the case of flavor SU(3) symmetry, whereas
the fourth one does the total results with the linear $m_s$
corrections taken into account. Note that for the transition magnetic
moments the relative signs are not important, because the
wavefunctions for the heavy baryons can include a phase
factor. Thus, the sign differences between different models do not
matter. So, we will compare the magnitudes of the transition magnetic
moments with those from other works. 
\begin{table}[htp]
  \global\long\def\arraystretch{1.3}
\caption{Numerical results of the transition magnetic moments for
the charmed baryons in units of $\mu_{N}$. They are compared with
those from the nonrelativistic quark
models~\cite{Johnson:1976mv,Choudhury:1976dp,Lichtenberg:1976fi} 
and Ref.~\cite{Franklin:1981rc} for $B_c$ corresponding to 
${\bf 6}_{1/2}$ and ${\bf 6}_{3/2}$, respectively.
The values from the modified bag model~\cite{Bernotas:2013eia},
light-cone QCD sum rules~\cite{Aliev:2001xr,Aliev:2009jt},
chiral perturbation theory~\cite{Wang:2018cre},
the chiral constituent quark model~\cite{Sharma:2010vv},  the 
Skyrme model (SM)~\cite{Oh:1991ws}, 
and the bound state approach with heavy quark
symmetries~\cite{Scholl:2003ip} are also listed. In the last column,
we list the results from a lattice QCD
simulation~\cite{Bahtiyar:2015sga, Bahtiyar:2016dom}.}  
\label{tab:1} 
  
\centering

\begin{tabular}{|c|cc|cccccccc|}
\hline 
$B_{c}\rightarrow B_{c}^{\prime}$  
& $\mu^{(0)}$$\left[\mu_{N}\right]$  
& $\mu^{(\text{total})}$ $\left[\mu_{N}\right]$  
  & \cite{Johnson:1976mv,Choudhury:1976dp,Lichtenberg:1976fi,
    Franklin:1981rc}  
& \cite{Bernotas:2013eia}  
& \cite{Aliev:2001xr,Aliev:2009jt}  
& \cite{Wang:2018cre}  
& \cite{Sharma:2010vv}  
& \cite{Oh:1991ws}  
& \cite{Scholl:2003ip}
& \cite{Bahtiyar:2015sga, Bahtiyar:2016dom}
\tabularnewline
\hline 
\hline 
$\Sigma_{c}^{+}\rightarrow\Lambda_{c}^{+}$  
& $1.24\pm0.05$  
& $1.54\pm0.06$  
& $1.63$  
& --  
& $-1.5\pm0.4$ 
& $-1.38\pm0.02$  
& $1.56$  
& $-1.67$  
& $-2.26$ 
&
\tabularnewline
$\Xi_{c}^{\prime+}\rightarrow\Xi_{c}^{+}$  
& $-1.24\pm0.05$  
& $-1.19\pm0.06$  
& $1.56$  
& --  
& --  
& $0.73(\mathrm{input})$  
& $1.30$  
& --  
& -- 
& $2.036$
\tabularnewline
$\Xi_{c}^{\prime0}\rightarrow\Xi_{c}^{0}$  
& $0$  
& $0.21\pm0.03$  
& $-0.07$  
& --  
& --  
& $0.22$  
& $-0.31$  
& --  
& -- 
& $0.039$
\tabularnewline
\hline 
$\Sigma_{c}^{\ast+}\rightarrow\Lambda_{c}^{+}$  
& $-1.76\pm0.08$  
& $-2.18\pm0.08$  
& $2.2$  
& $1.70$  
& $2.00\pm0.53$  
& $2.00$  
& $2.40$  
& --  
& -- 
&
\tabularnewline
$\Xi_{c}^{\ast+}\rightarrow\Xi_{c}^{+}$  
& $1.76\pm0.08$  
& $1.69\pm0.08$  
& $2.03$  
& $1.50$  
& $1.93\pm0.72$  
& $1.05$  
& $2.08$  
& --  
& -- 
&
\tabularnewline
$\Xi_{c}^{\ast0}\rightarrow\Xi_{c}^{0}$  
& $0$  
& $-0.29\pm0.04$  
& $-0.33$  
& $-0.22$  
& $0.22\pm0.07$  
& $-0.31$  
& $-0.50$  
& --  
& -- 
&
\tabularnewline
\hline 
$\Sigma_{c}^{\ast++}\rightarrow\Sigma_{c}^{++}$  
& $1.42\pm0.07$  
& $1.52\pm0.07$  
& $1.39$  
& $0.91$  
& $1.33\pm0.38$
& $1.07\pm0.23$  
& $-1.37$  
& --  
& --  
&
\tabularnewline
$\Sigma_{c}^{\ast+}\rightarrow\Sigma_{c}^{+}$  
& $0.35\pm0.02$  
& $0.33\pm0.02$  
& $0.07$  
& $-0.06$  
& $0.57\pm0.09$  
& $0.19\pm0.06$  
& $-0.003$  
& --  
& --  
&
\tabularnewline
$\Sigma_{c}^{\ast0}\rightarrow\Sigma_{c}^{0}$  
& $-0.71\pm0.03$  
& $-0.87\pm0.03$  
& $-1.24$  
& $-1.03$  
& $0.24\pm0.05$  
& $-0.69\pm0.1$  
& $1.48$  
& --  
& --  
&
\tabularnewline
$\Xi_{c}^{\ast+}\rightarrow\Xi_{c}^{\prime+}$  
& $0.35\pm0.02$  
& $0.43\pm0.02$  
& $0.09$  
& $-0.09$  
& --  
& $0.23\pm0.06$  
& $-0.23$  
& --  
& --  
&
\tabularnewline
$\Xi_{c}^{\ast0}\rightarrow\Xi_{c}^{\prime0}$  
& $-0.71\pm0.03$  
& $-0.74\pm0.03$  
& $-1.07$  
& $-0.92$  
& --  
& $-0.59\pm0.12$  
& $1.24$  
& --  
& --  
&
\tabularnewline
$\Omega_{c}^{\ast0}\rightarrow\Omega_{c}^{0}$  
& $-0.71\pm0.03$  
& $-0.60\pm0.04$  
& $-0.94$  
& $-0.84$  
& --  
& $-0.49\pm0.14$  
& $0.96$  
& --  
& --  
& $0.658$
\tabularnewline
\hline 
\end{tabular}
\end{table}

As explained in the previous Section, the effects of flavor SU(3)
symmetry breaking come from both the collective operators with
$\overline{w}_{4,5,6}$ and the collective wavefunctions. The total
results listed in Table~\ref{tab:1} contain both the contributions,
though we do not show them separately. In general, the effects from
the collective operators are dominant over those from the
wavefunction corrections. However, the wavefunction corrections are
not negligible specifically in the transitions    
$\Sigma_c^+\to \Lambda_c^+\gamma$, $\Xi_c'^{0}\to \Xi_c^0 \gamma $,
$\Sigma_c^{*+}\to \Lambda_c^+\gamma$, and $\Xi_c^{*0}\to \Xi_c^0
\gamma$. As shown 
in Eqs.~\eqref{eq:61_3bar_2} and \eqref{eq:63_3bar_2}, the transition
magnetic moments for the decays $\Xi_c'\to \Xi_c \gamma$ and $\Xi_c^*\to
\Xi_c \gamma$ are proportional to the corresponding charge $\mathcal
Q$. Thus, the transition magnetic moments for the radiative decays of
the neutral $\Xi_c'$ and $\Xi_c^*$ baryons vanish. This means that the
linear $m_s$ corrections take over the leading-order role. In this
case, the effects from the wavefunctions are 
still important. They contribute to the results of $\mu[\Xi_c'^{0}\to
\Xi_c^0]$ and $\mu[\Xi_c^{*0}\to \Xi_c^0]$ by about 30~\%.
It is interesting to see that the effects of SU(3) symmetry breaking
contribute to the transitions of the $\Sigma_c^+ \to \Lambda_c^+\gamma$,
$\Sigma_c^{*+} \to \Lambda_c^+\gamma$, $\Sigma_c^{*0}\to \Sigma_c^0
\gamma$, $\Xi_c^{*+}\to \Xi_c'^{+}\gamma$, and $\Omega_c^{*0}\to
\Omega_c^0 \gamma$ by about $20~\%$, whereas their contributions are
rather small to other decay channels (about $6~\%$). Moreover, the
effects of SU(3) symmetry breaking add to the magnitudes of the
transition magnetic moments in certain channels but reduce those of
the $\Xi_c'^+\to \Xi_c^+\gamma$, $\Xi_c^{*+}\to \Xi_c^{+}\gamma$,
$\Sigma_c^{*+}\to \Sigma_c^+\gamma$, and $\Omega_c^{*0}\to \Omega_c^0
\gamma$ channels. We can analyze these 
interesting results by examining the operator parts of SU(3) symmetry 
breaking shown in Eqs.~\eqref{eq:63_3bar_2} and~\eqref{eq:63_61_c}.
For example, the linear $m_s$ corrections from the operator for the
$\Xi_c'^+\to \Xi_c^+\gamma$ decay is proportional to
$[(\mathcal{Q}-1)\overline{w}_4 + 2(\mathcal{Q}-1) \overline{w}_5
-(2\mathcal{Q}-1) \overline{w}_6]$. Since $\Xi_c'^+$ has a positive
charge, the $m_s$ correction is only proportional to
$-\overline{w}_6$, which gives a positive contribution. On the other  
hand, the SU(3) symmetric part yields the negative result, so that the
magnitude of the transition magnetic moment of the $\Xi_c'^+\to
\Xi_c^+\gamma$ is diminished by the $m_s$ corrections. One can
understand other decay channels in the same way. Note that a similar
tendency was  found in the case of the $\Xi^{*0} \to \Xi^0 \gamma$ 
decay~\cite{Kim:2005gz}.   

We also compare the present results with those from the 
NRQM~\cite{Johnson:1976mv,Choudhury:1976dp,Lichtenberg:1976fi}, the
modified bag model~\cite{Bernotas:2013eia}, the chiral constituent
quark model~\cite{Sharma:2010vv}, the light-cone QCD sum
rules~\cite{Aliev:2001xr,Aliev:2009jt}, and the Skyrme models with
bound state approaches~\cite{Oh:1991ws,Scholl:2003ip}.  The present
are comparable with those from all other works. In particular, it is
of great interest to compare the present results with those of a
recent simulation of lattice QCD~\cite{Bahtiyar:2015sga,
  Bahtiyar:2016dom}, which are listed in the last column.  Note that
in Refs.~\cite{Bahtiyar:2015sga,   Bahtiyar:2016dom} a value of the
pion mass $m_\pi=156$ MeV was used, which is quite close to the
experimental value. Apart from the signs, the present results are in
qualitative agreement with the lattice data. The result of the
$\Omega_c^{*0} \to \Omega_c \gamma$ decay is in good agreement with
that of Ref.~\cite{Bahtiyar:2015sga}, compared to those from other
models. 

\begin{table}[H]
\global\long\def\arraystretch{1.3}
 \caption{Numerical results of the transition magnetic moments for
the bottom baryons in units of $\mu_{N}$. They are compared with
those from the nonrelativistic quark model from
Refs.~\cite{Johnson:1976mv,Choudhury:1976dp,Lichtenberg:1976fi}
and Ref.~\cite{Franklin:1981rc} for $B_c$ corresponding to 
${\bf 6}_{1/2}$ and ${\bf 6}_{3/2}$, respectively.
The values from modified bag model~\cite{Bernotas:2013eia},
light-cone QCD sum rule \cite{Aliev:2001xr,Aliev:2009jt},
chiral perturbation theory \cite{Wang:2018cre},
Skyrme model (SM) \cite{Oh:1991ws}, 
and bound state approach \cite{Scholl:2003ip} are also listed.}
\label{tab:2}
\centering
\begin{tabular}{|c|cc|cccccc|}
\hline 
$B_{b}\rightarrow B_{b}^{\prime}$  
& $\mu^{(0)}$$\left[\mu_{N}\right]$  
& $\mu^{(\text{total})}$$\left[\mu_{N}\right]$  
& \cite{Johnson:1976mv,Choudhury:1976dp,Lichtenberg:1976fi,Franklin:1981rc}  
& \cite{Bernotas:2013eia}  
& \cite{Aliev:2001xr,Aliev:2009jt}  
& \cite{Wang:2018cre}  
& \cite{Scholl:2003ip}  
& \cite{Oh:1991ws} 
\tabularnewline
\hline 
\hline 
$\Sigma_{b}^{0}\rightarrow\Lambda_{b}^{0}$  
& $-1.24\pm0.05$  
& $-1.54\pm0.06$  
& --  
& --  
& --  
& $-1.37$  
& $-2.24$  
& $-1.54$ 
\tabularnewline
$\Xi_{b}^{\prime0}\rightarrow\Xi_{b}^{0}$  
& $1.24\pm0.05$  
& $1.19\pm0.06$  
& --  
& --  
& --  
& $-0.75$  
& --  
& -- 
\tabularnewline
$\Xi_{b}^{\prime-}\rightarrow\Xi_{b}^{-}$  
& $0$  
& $-0.21\pm0.03$  
& --  
& --  
& --  
& $0.21$  
& --  
& -- 
\tabularnewline
\hline 
$\Sigma_{b}^{\ast0}\rightarrow\Lambda_{b}^{0}$  
& $-1.76\pm0.08$  
& $-2.18\pm0.08$  
& $2.28$  
& $1.49$  
& $1.52\pm0.58$  
& $1.96$  
& --  
& -- 
\tabularnewline
$\Xi_{b}^{\ast0}\rightarrow\Xi_{b}^{0}$  
& $1.76\pm0.08$  
& $1.69\pm0.08$  
& $2.03$  
& $1.32$  
& $1.71\pm0.60$  
& $1.06$  
& --  
& -- 
\tabularnewline
$\Xi_{b}^{\ast-}\rightarrow\Xi_{b}^{-}$  
& $0$  
& $-0.29\pm0.04$  
& $-0.26$  
& $-0.14$  
& $0.18\pm0.06$  
& $-0.30$  
& --  
& -- 
\tabularnewline
\hline 
$\Sigma_{b}^{\ast+}\rightarrow\Sigma_{b}^{+}$  
& $-1.42\pm0.07$  
& $-1.52\pm0.07$  
& $1.81$  
& $1.19$  
& $0.83\pm0.25$  
& $1.17\pm0.22$  
& --  
& -- 
\tabularnewline
$\Sigma_{b}^{\ast0}\rightarrow\Sigma_{b}^{0}$  
& $-0.35\pm0.02$  
& $-0.33\pm0.02$  
& $0.49$  
& $0.35$  
& $0.20\pm0.08$  
& $0.30\pm0.06$  
& --  
& -- 
\tabularnewline
$\Sigma_{b}^{\ast-}\rightarrow\Sigma_{b}^{-}$  
& $0.71\pm0.03$  
& $0.87\pm0.03$  
& $-0.82$  
& $-0.50$  
& $0.42\pm0.14$  
& $-0.58\pm0.1$  
& --  
& -- 
\tabularnewline
$\Xi_{b}^{\ast0}\rightarrow\Xi_{b}^{\prime0}$  
& $-0.35\pm0.02$  
& $-0.43\pm0.02$  
& $0.61$  
& $0.39$  
& --  
& $0.33\pm0.06$  
& --  
& -- 
\tabularnewline
$\Xi_{b}^{\ast-}\rightarrow\Xi_{b}^{\prime-}$  
& $0.71\pm0.03$  
& $0.74\pm0.03$  
& $-0.66$  
& $-0.42$  
& --  
& $-0.49\pm0.1$  
& --  
& -- 
\tabularnewline
$\Omega_{b}^{\ast-}\rightarrow\Omega_{b}^{-}$  
& $0.71\pm0.03$  
& $0.60\pm0.04$  
& $-0.52$  
& $-0.34$  
& --  
& $-0.38\pm0.13$  
& --  
& -- 
\tabularnewline
\hline 
\end{tabular}
\end{table}
Table~\ref{tab:2} lists the results of the transition magnetic moments
of the bottom baryons. In the present pion mean-field approach, we
take the limit of the infinitely heavy quark mass ($m_Q\to
\infty$). Thus, the results are in fact the same as those of the
charmed baryons. Only the signs of the results are different, because
charge of the bottom quark $\mathcal{Q}_b=-1/3$ is different from that of the
charm quark $\mathcal{Q}_c=2/3$. Nevertheless, the results are quite
comparable with those of other works. In particular, we find that the
present result of the $\Sigma_b^0\to \Lambda_b^0\gamma$ is almost the
same as that from the Skyrme model~\cite{Oh:1991ws}.  

\begin{table}[H]
\global\long\def\arraystretch{1.3}
\caption{Numerical results of the radiative decay rates for the charmed
baryons in units of $\mathrm{keV}$. The results are compared with
those from the modified bag model \cite{Bernotas:2013eia}, light-cone
QCD sum rule~\cite{Aliev:2009jt}, chiral perturbation
theory~\cite{Banuls:1999br, Wang:2018cre},  and lattice
QCD~\cite{Bahtiyar:2015sga,Bahtiyar:2016dom}. }
\label{tab:3}
\centering
\begin{tabular}{|c|cc|ccccc|}
\hline 
$B_{c}\rightarrow B_{c}^{\prime}\;\gamma$  
& $\Gamma_{\gamma}^{\left(0\right)}\left[\mathrm{keV}\right]$  
& $\Gamma_{\gamma}^{\left(\mathrm{total}\right)}\left[\mathrm{keV}\right]$  
& \cite{Bernotas:2013eia}  
& \cite{Aliev:2009jt}  
& \cite{Banuls:1999br}
& \cite{Wang:2018cre}  
& \cite{Bahtiyar:2015sga,Bahtiyar:2016dom} 
\tabularnewline
\hline 
\hline 
$\Sigma_{c}^{+}\rightarrow\Lambda_{c}^{+}\;\gamma$  
& $8.32\pm0.73$  
& $12.82\pm0.95$  
& --  
& --   
& --  
& $65.6\pm2$  
& -- 
\tabularnewline
$\Xi_{c}^{\prime+}\rightarrow\Xi_{c}^{+}\;\gamma$  
& $2.18\pm0.20$  
& $2.02\pm0.20$  
& --  
& --   
& --  
& $5.43\pm0.33$  
& $5.468$ 
\tabularnewline
$\Xi_{c}^{\prime0}\rightarrow\Xi_{c}^{0}\;\gamma$  
& $0$  
& $0.06\pm0.01$  
& --  
& --   
& $1.2\pm0.7$  
& $0.46$  
& $0.002$ 
\tabularnewline
\hline 
$\Sigma_{c}^{\ast+}\rightarrow\Lambda_{c}^{+}\;\gamma$  
& $23.04\pm2.12$  
& $35.49\pm2.81$  
& $21.61$  
& $29.90$   
& --  
& $161.6\pm5$  
& -- 
\tabularnewline
$\Xi_{c}^{\ast+}\rightarrow\Xi_{c}^{+}\;\gamma$  
& $9.34\pm0.82$  
& $8.66\pm0.81$  
& $6.82$  
& $11.29$   
& --  
& $21.6\pm1$  
& -- 
\tabularnewline
$\Xi_{c}^{\ast0}\rightarrow\Xi_{c}^{0}\;\gamma$  
& $0$  
& $0.25\pm0.06$  
& $0.14$  
& $0.14$   
& $5.1\pm2.7$  
& $1.84$  
& -- 
\tabularnewline
\hline 
$\Sigma_{c}^{\ast++}\rightarrow\Sigma_{c}^{++}\;\gamma$  
& $0.31\pm0.03$  
& $0.36\pm0.03$  
& $0.13$  
& $0.28$  
& --   
& $1.20\pm0.6$  
& -- 
\tabularnewline
$\Sigma_{c}^{\ast+}\rightarrow\Sigma_{c}^{+}\;\gamma$  
& $0.02\pm0.003$  
& $0.02\pm0.003$  
& $0.001$  
& $0.05$   
& --  
& $0.04\pm0.03$  
& -- 
\tabularnewline
$\Sigma_{c}^{\ast0}\rightarrow\Sigma_{c}^{0}\;\gamma$  
& $0.08\pm0.01$  
& $0.12\pm0.01$  
& $0.17$  
& $0.01$   
& --  
& $0.49\pm0.1$  
& -- 
\tabularnewline
$\Xi_{c}^{\ast+}\rightarrow\Xi_{c}^{\prime+}\;\gamma$  
& $0.02\pm0.002$  
& $0.03\pm0.003$  
& $0.001$  
& --   
& --  
& $0.07\pm0.03$  
& -- 
\tabularnewline
$\Xi_{c}^{\ast0}\rightarrow\Xi_{c}^{\prime0}\;\gamma$  
& $0.08\pm0.01$  
& $0.09\pm0.01$  
& $0.14$  
& --  
& --   
& $0.42\pm0.16$  
& -- 
\tabularnewline
$\Omega_{c}^{\ast0}\rightarrow\Omega_{c}^{0}\;\gamma$  
& $0.09\pm0.01$  
& $0.06\pm0.01$  
& $0.12$  
& --   
& --  
& $0.32\pm0.20$  
& $0.074$ 
\tabularnewline
\hline 
\end{tabular}
\end{table}
It is straightforward to compute the radiative decay rates of the
heavy baryons by using Eqs.~\eqref{eq:parwidth1} and
\eqref{eq:parwidth2}, once we have known the corresponding transition
magnetic moments. The results for the charmed baryons are presented in
Table~\ref{tab:3}. They are quite similar to those from
the modified bag model~\cite{Bernotas:2013eia} and light-cone sum
rule~\cite{Aliev:2009jt}, the  whereas they are somewhat deviated from 
those of Ref.~\cite{Wang:2018cre} even though the results of the
transition magnetic moments are comparable each other. On the other
hand, lattice QCD, $\chi$PT of Ref.~\cite{Banuls:1999br} and the quark
models used the same formulae as in the present work. 
If we recalculate the radiative decay
rates, using the results of the transition magnetic moments from
Ref.~\cite{Wang:2018cre}, we find that 
\begin{align}
\Gamma_{\Sigma_c^+\to \Lambda_c^+\gamma}&=11.06 \,\mathrm{keV},\;\;\;
\Gamma_{\Xi_c'^+\to \Xi_c^+\gamma}=0.79 \,\mathrm{keV},\;\;\;
\Gamma_{\Xi_c'^+\to \Xi_c^+\gamma}=0.07 \,\mathrm{keV},\cr
\Gamma_{\Sigma_c^{*+} \to \Lambda_c^+\gamma}&=18.17
 \,\mathrm{keV},\;\;\;
\Gamma_{\Xi_c^{*+} \to \Xi_c^+\gamma}=2.08 \,\mathrm{keV},\;\;\;
\Gamma_{\Xi_c^{*0} \to \Xi_c^0\gamma}=0.18 \,\mathrm{keV}, 
  \label{eq:xpt}
\end{align}
which are quite close to the present ones. 
Since Ref.~\cite{Banuls:1999br} presented only the results of the
radiative decay widths from $\chi$PT, we compare the present ones with
them. Those of Ref.~\cite{Banuls:1999br} are larger than those
of all other models as well as the present results. 
In the last column of Table~\ref{tab:3}, the results from the
calculation of lattice QCD are
listed~\cite{Bahtiyar:2015sga,Bahtiyar:2016dom}, which are also
comparable to the present ones.   

\begin{table}[H]
\global\long\def\arraystretch{1.3}
 \caption{Numerical results of the radiative decay rates for the bottom
baryons in units of $\mathrm{keV}$. The results are compared with
those from the modified bag model~\cite{Bernotas:2013eia}, light-cone
QCD sum rule~\cite{Aliev:2009jt}, and chiral perturbation theory
\cite{Banuls:1999br, Wang:2018cre}. }
\label{tab:4} 
\centering
\begin{tabular}{|c|cc|cccc|}
\hline 
$B_{b}\rightarrow B_{b}^{\prime}\;\gamma$  
& $\Gamma_{\gamma}^{\left(0\right)}\left[\mathrm{keV}\right]$  
& $\Gamma_{\gamma}^{\left(\mathrm{total}\right)}\left[\mathrm{keV}\right]$  
& \cite{Bernotas:2013eia}  
& \cite{Aliev:2009jt}  
& \cite{Banuls:1999br}
& \cite{Wang:2018cre} 
\tabularnewline
\hline 
\hline 
$\Sigma_{b}^{0}\rightarrow\Lambda_{b}^{0}\;\gamma$  
& $2.4\pm0.2$  
& $3.7\pm0.3$  
& --  
& --  
& --
& $108.0\pm4$
\tabularnewline
$\Xi_{b}^{\prime0}\rightarrow\Xi_{b}^{0}\;\gamma$  
& $0.93\pm0.08$  
& $0.87\pm0.08$  
& --  
& --   
& --
& $13.0\pm0.8$
\tabularnewline
$\Xi_{b}^{\prime-}\rightarrow\Xi_{b}^{-}\;\gamma$  
& $0$  
& $0.02\pm0.01$  
& --  
& --   
& $3.1\pm1.8$
& $1.0$
\tabularnewline
\hline 
$\Sigma_{b}^{\ast0}\rightarrow\Lambda_{b}^{0}\;\gamma$  
& $3.3\pm0.3$  
& $5.1\pm0.4$  
& $2.38$  
& $2.48$   
& --
& $142.1\pm5$
\tabularnewline
$\Xi_{b}^{\ast0}\rightarrow\Xi_{b}^{0}\;\gamma$  
& $1.3\pm0.1$  
& $1.2\pm0.1$  
& $0.72$  
& $1.20$   
& --
& $17.2\pm0.1$
\tabularnewline
$\Xi_{b}^{\ast-}\rightarrow\Xi_{b}^{-}\;\gamma$  
& $0$  
& $0.03\pm0.01$  
& $0.01$  
& $0.01$   
& $4.2\pm2.4$
& $1.4$
\tabularnewline
\hline 
$\Sigma_{b}^{\ast+}\rightarrow\Sigma_{b}^{+}\;\gamma$  
& $0.0020$  
& $0.0022$  
& $0.0014$  
& $0.0007$  
& -- 
& $50\pm20$
\tabularnewline
$\Sigma_{b}^{\ast0}\rightarrow\Sigma_{b}^{0}\;\gamma$  
& $0.0001$  
& $0.0001$  
& $0.0001$  
& $0.00004$   
& --
& $3.0\pm1$
\tabularnewline
$\Sigma_{b}^{\ast-}\rightarrow\Sigma_{b}^{-}\;\gamma$  
& $0.0004$  
& $0.0010$  
& $0.0002$  
& $0.0001$   
& --
& $10.3\pm4$
\tabularnewline
$\Xi_{b}^{\ast0}\rightarrow\Xi_{b}^{\prime0}\;\gamma$  
& $0.00004$  
& $0.0001$  
& $0.00005$  
& --   
& --
& $1.5\pm0.5$
\tabularnewline
$\Xi_{b}^{\ast-}\rightarrow\Xi_{b}^{\prime-}\;\gamma$  
& $0.0004$  
& $0.0005$  
& $0.0001$  
& --   
& --
& $8.2\pm4$
\tabularnewline
$\Omega_{b}^{\ast-}\rightarrow\Omega_{b}^{-}\;\gamma$  
& $0.006\pm0.001$  
& $0.004\pm0.001$  
& $0.0013$  
& --   
& --
& $30.6\pm26$
\tabularnewline
\hline 
\end{tabular}
\end{table}
In Table~\ref{tab:4}, we list the results of the radiative decay
rates for the bottom baryons. Again, they are comparable with those
from the other works. By the same reason, the results from
$\chi$PT~\cite{Wang:2018cre} are different from the present ones but
the recalculated results with Eqs.~\eqref{eq:parwidth1} and
\eqref{eq:parwidth1} are quite comparable with the present ones. 

\section{Summary and Conclusions}
In the present work, we have investigated the transition magnetic
moments of the lowest-lying heavy baryon sextets within a pion
mean-field approach or a ``\emph{model-independent} chiral
quark-soliton model. Since all the dynamical parameters for the
magnetic moments of the singly-heavy baryons were fixed in the light
baryon sector, we were able to compute the transition magnetic moments
of the baryon sextet without any additional parameters introduced. We
also derived the various relations among the transition magnetic
moments as in the case of the magnetic moments of the heavy
baryons. In addition, we found the relations that arise from the
$U$-spin symmetry. The results of the transition magnetic moments for
the charmed and bottom baryons were compared with those from other
works. In particular, the present results are in good agreement with
those from a simulation of lattice QCD. We obtained the radiative
decay rates of the heavy baryons and compared the results with those
from other works. While they are quite comparable with the results
from the modified bag model, the light-cone QCD sum rule, and the
simulation of lattice QCD, those from chiral perturbation theory seem
different from the present ones. The reason arises from the fact that
different formulae for the radiative decay rates were used. Using the
same formulae, we found that the present results are also in good
agreement with those from chiral perturbation theory.

Information on the transition magnetic moments may provide the vector
meson coupling to the heavy baryons. Though the experimental data on
them are still absent, theoretical investigations may shed lights on
how the vector mesons can be coupled to the heavy baryons. In fact,
this will provide essential information on the hadronic description of
heavy hadron productions. The corresponding study is under way. 

\section*{Acknowledgments}
The authors are grateful to J.-Y. Kim for valuable discussions. 
The present work was supported by Basic Science Research Program
through the National Research Foundation of Korea funded by the
Ministry of Education, Science and Technology
(NRF-2019R1A2C1010443 (Gh.-S. Y.), 2018R1A2B2001752, and
2018R1A5A1025563 (H.-Ch.K.)).

\end{document}